\documentclass{article}
\usepackage[utf8]{inputenc}
\usepackage{geometry}
 \usepackage{graphicx}
 \usepackage{titling}
 \usepackage{color}

\definecolor{title}{rgb}{0.5,0.1,0.5}
\definecolor{section}{rgb}{0.4,0,0.4}
\definecolor{subsection}{rgb}{0.4,0.3,0.5}

 \title{\textbf{\huge \color{title} Advocacy for Physics and for Physicists} \\ \bigskip \textit{Results of an Informal Survey of \\ American Physical Society Members in 2025}}

\author{Michael B. Bennett, for the Physics Advocacy Collaboration}
\date{June 2025}
 
 \usepackage{fancyhdr}
\fancypagestyle{plain}{%  the preset of fancyhdr 
    \fancyhf{} % clear all header and footer fields
    \fancyfoot[R]{}
    \fancyfoot[L]{\thedate}
    \fancyhead[L]{Report}
    \fancyhead[R]{\theauthor}
}
\makeatletter
\def\@maketitle{%
  \newpage
  \null
  \vskip 1em%
  \begin{center}%
  \let \footnote \thanks
    {\LARGE \@title \par}%
    \vskip 1em%
  \end{center}%
  \par
  \vskip 1em}
\makeatother

\usepackage{lipsum}  
\usepackage{cmbright}
\usepackage{parskip}
\fontfamily{lato}
\usepackage{wrapfig}
\usepackage{booktabs}
\usepackage{appendix}
\usepackage{pdfpages}
\usepackage{url}
\usepackage{hyperref}
\hypersetup{colorlinks=true, citecolor=subsection, linkcolor=section, urlcolor=section}
\usepackage[style=numeric-comp,sorting=none]{biblatex}
    \bibliography{report}

\newcommand{\callout}[1]{\textbf{\color{subsection}#1}}

\usepackage{xcolor}
\usepackage{sectsty}

\sectionfont{\color{section}}
\subsectionfont{\color{subsection}}
\subsubsectionfont{\color{subsection}}

\begin{document}

\maketitle

\bigskip

\hrule

\section*{Executive Summary}
\label{sec:exec_summary}
This report was written in June 2025, during the first year of the second Trump administration and in the context of several actions by the Executive Branch of the United States Government to reduce the scope of U.S. science activity, including through a Fiscal Year 2026 Presidential Budget Request that would dramatically cut funding to national agencies that make research possible.  In the shadow of these actions, professional scientific societies in the U.S. have taken a variety of different responses.  

Our collaboration represents a diverse group of physicists, including faculty, staff, postdocs, etc., all members of the American Physical Society (APS).  As APS members, we are particularly interested in the extent to which APS activity in the wake of the administration's efforts to scale back science actually serves the needs of APS members both in the short and long term, especially given APS's stated commitment to values against which the current administration has explicitly set itself.  To investigate these questions of alignment between organizational values, members needs, and Society action, we conducted an informal survey of APS members inquiring about their experiences with APS communication since the 2025 inauguration, their needs in the present moment, and what types of actions would be helpful to them.  

This report provides detail on the context for the creation and implementation of the survey, as well as articulating results and some common themes found in responses.  We also include some suggestions of considerations for APS and other scientific professional societies who want to do right by members in a perilous political climate.  This report is not meant to present as academic research, and has not been submitted for peer review; rather, by drawing on the research expertise of multiple collaboration members, we have endeavored to create an authentic snapshot of APS membership in 2025, and we use this report as a starting point for ongoing conversations about how APS can better serve its members.

Broadly, our informal survey of APS members revealed that a majority of respondents perceive that the Society supports them, cares about their needs as physicists, and is communicating reliably and transparently.  \callout{We celebrate these results, but we note as well that a sizeable proportion of respondents -- up to 30\% -- harbor concerns about the organization's actions, its prioritization of member needs over corporate interests, or its willingness to listen to members.  In addition, a strong majority -- over two-thirds of respondents -- articulated a desire for more concrete and personal support from APS.}  While we acknowledge that the Society cannot be all things to all people, the results of our work lead us to the concern that there may indeed be a disconnect between the APS envisioned by those at the top of the organization and those who fill its ranks.  \callout{We therefore exhort APS first and foremost to demonstrate solidarity with its members, many of whom are being directly harmed as a result of the current administrative climate, by listening to them and giving them a platform to shape the organization whose body they comprise.}

\bigskip
\smallskip
\smallskip

\hrule

\newpage

\section{Social and Professional Context}
\label{sec:context}

Since the January 2025 inauguration of United States President Trump's second term, the administration has taken multiple unambiguous steps to control the scope, scale, reach, and focus of U.S. scientific activity.  As early as January 21, 2025, one day after the inauguration, with the administration's pausing of public communication for the National Institute of Health (NIH), the Center for Disease Control (CDC), and the Food and Drug Administration (FDA) \cite{NIH_pause_2025}, and as recently as June 10, 2025, when the administration's Health Secretary, Robert F. Kennedy Jr., unilaterally fired all 17 members of the Advisory Committee for Immunization Practices \cite{kennedy_fire_2025}, the administration has signalled repeatedly its intent to control not only how science is performed and communicated, but who is allowed to do it.  Of greatest import for most physicists, the Presidential Budget Request (PBR) for the 2026 fiscal year slashes funding for the National Science Foundation (NSF) by nearly 70\% \cite{NSF_Budget}.  The NSF, which as of the publication of this report has been ordered out of its own building in Alexandria, VA \cite{NSF_moved} funds, along with five other government institutions, up to 90\% basic research in the country \cite{NSF_funding}, so a cut this drastic already represents a substantial threat to physics as a professional field, but even more alarming is the fact that these cuts are not being proposed equally across the board.

Rather, multiple analyses of the cuts have revealed that they appear targeted to systematically destroy STEM education research, DEI work, and work by women and people of color \cite{NYT_STEM_2025,NYT_Budget_2025,Propublica_Women_2025}, with more diverse agencies being preferentially targeted by reductions.  Of course, the immediate impact of these reductions is the loss of crucial talent in the government, at universities, and throughout academia, but the longer-term impacts -- the ``chilling'' of research efforts, a catastrophic loss of future talent, widespread economic loss, and America's leadership of the scientific world \cite{Duke_Chilling_2025, AIP_physicsgrads_2025,Forbes_Cost_2025,Economist_impacts_2025,Lancet_impacts_2025} -- have only begun to be felt and are practically impossible to quantify at this point.

Resistance to these cuts and reductions has taken the form of large- and small-scale protests, mass resignations of high-level federal workers, and a number of lawsuits joined by state Attorneys General as well as other parties to challenge the legality of the administration's attempts to override congressional authority \cite{Waging_Resistance_2025,Science_Panch_2025,NPR_Fullbright_2025,AP_Tracking_2025}.  The organizational scientific response to the administration's attacks has been varied; while some professional science organizations like the American Geophysical Union, have been clear and unambiguous both in their opposition to the administration's efforts \cite{AGU_StateoftheUnion_2025} and in their support of the people that make up the scientific enterprise \cite{AGU_Community_2025}, many have remained somewhat circumspect in their communications.  For example, a March 21 comment from the chair of the American Chemical Society (ACS) does not acknowledge or name the administration or any of its actions in its discussion of ACS activity \cite{ACS_comment_2025}; this comment drew ire and further discussion from ACS members dissatisfied with perceived complicity through silence \cite{ACS_lettereditor_2025}.

\subsection{Motivation}
\label{sec:motivation}

The impetus for this report began following a member-wide mailer from the APS CEO, Jonathan Bagger, on February 6 2025.  The letter affirms that ``[APS's] mission is to advance physics by fostering a vibrant, inclusive, and global community dedicated to science and society'' and that ``Our core values guide our work -- both now and in the future.'' \cite{APS_mailer1_2025}.  The letter acknowledges that federal freezing of research support ``\ldots impacts careers and harms U.S. competitiveness,'' but stops short of attributing the actions to the administration or articulating or linking to any specific values or mission.  Multiple members of the APS Committee for Public Engagement brought to the attention of APS senior leadership (a list of head APS personnel can be found at Ref. \cite{propublica_APS_profile}) the concern that communications like this one did not take strong enough stance in favor of APS's most vulnerable members, those being most directly impacted by federal actions.  

Following this conversation, our collaboration attended the APS Global Physics summit ``Protecting Science Town Hall'' on Thursday, March 20 and proposed a joint staff-member task force to accomplish three goals: {\color{section}(i)} identify member needs and interests in the current political climate, {\color{section}(ii)} improve communication to members on the ways in which APS actions  potentially already meet their needs, and {\color{section}(iii)} identify new potential initiatives or actions by which APS could more nearly meet the needs of its members.  We encourage readers who missed one or both Town Halls to listen to the recorded audio, still available as of publication of this report at Ref. \cite{APS_GPS_Video}.  The first objective of the task force was envisioned to be the creation and distribution of a member-focused survey of needs.

As part of its ``Standing Up for Science'' lobbying campaign, APS had by this point already sent out an ``Impacts of Federal Science Funding'' survey (still up as of the publication of this report, Ref. \cite{APS_Surveylink}); although that survey does collect member impacts, it is designed to help identify individuals to participate in the campaign and does not ask about impacts to APS members other than loss of federal funding.  It also requires respondents to identify themselves and give contact information, which can mean that vulnerable parties are less likely to respond for fear of the information being shared with external threatening parties (in fact, part of the impetus for designing a dedicated member needs survey was having heard from multiple APS members from marginalized groups that they specifically felt unsafe sharing their needs with APS given the requirement to self-identify).  As of publication, APS has also distributed a member-wide ``pulse check'' seemingly designed to link respondents to current APS activity, but again not well-designed for the purposes of recording members actual needs. \callout{Thus, we perceive that the organization is unfortunately largely ignorant of the needs of its members right now.  Both the organization and its members deserve to have that conversation -- especially since member needs and priorities have likely shifted since the inauguration.}  

The joint task force proposal at the GPS Town Hall was met with both enthusiastic support from audience members and an explicit affirmation of agreement from President Doyle: ``I'd love to have a conversation about this, especially the communications aspect, as you say.''  Following that response, we set up a line of communication with APS leadership and with multiple APS unit leaders.  Unfortunately, we were unable to garner a response from the three men who ran the town halls, the APS President, CEO, or Chief External Affairs Officer (CEAO). After a meeting with unit leaders and with members of APS Communications staff, we were unable to reach terms of collaboration with APS leadership.  Thus, we went ahead with an informal survey, by APS members, for APS members. This report is the culmination of that effort.

\section{Survey Design and Implementation}

The survey has been included at the end of this document as Appendix \ref{app:survey}.  We shared the survey with APS leadership before distribution and received input from multiple APS units in its design.  Broadly, the survey contained three banks of questions, all optional and all designed to investigate specific elements of APS members' experiences in the current political environment:
\begin{enumerate}
    \item Bank 1 included questions on respondents' awareness of and response to APS communications, such as the mass member mailers.
    \item Bank 2 included questions on respondents' immediate needs and the impacts of federal actions on their professional lives.
    \item Bank 3 included demographic questions through which respondents could identify if desired according to their career stage, role, etc.
\end{enumerate}

The survey included both four-point Likert-style question grids (ranging from a ``strongly disagree'' or similar option to a ``strongly agree'' or similar option, with no ``ambivalent'' option) and open-ended short response questions, in order to collect slightly richer data without unduly burdening respondents.  Because we were unable to collaborate with APS staff on survey distribution, we engaged in a grassroots distribution campaign utilizing APS Engage \cite{APS_Engage}, the organization's community platform.  Each APS unit has its own Engage hub; we distributed the survey on the community hubs for the Topical Group for Physics Education Research (multiple collaboration members are PER researchers), the Forum for Physics and Society (FPS), the Forum for Outreach and Engaging the Public (FOEP), the Forum for Early Career Scientists (FECS), and the Forum on Education (FEd).  We did not distribute directly to any APS Division platforms even though multiple collaboration members are of course members of different Divisions; the focus on Forums as the main avenue of distribution was because Forums are interest-driven and cut across different APS demographics and divisions.

We utilized the Fillout platform to build and implement the survey.  Fillout aggregates responses and summarizes them; since this effort was intended to provide an immediate snapshot of APS members in a certain point in time, very little additional analysis was required.  Most of what we present here is drawn directly from Fillout's aggregation summary; additional analysis consisted of ``gating'' the data on certain answers to see if common themes arose between respondents who answered certain ways on certain items (e.g., gating on early-career respondents).

Participation in the survey was completely confidential -- while we did include an option for respondents to share an email if they wanted to stay connected to our collaboration, that survey item linked to a separate email collection form, making it impossible for us to link email addresses with specific survey responses.  In this report, no individual responses will be shared, only themes that emerged from the data after analysis.  Participants could also share their responses while opting out of having their data included in the analysis for this report, essentially removing them from the data pool we share here.  Of 101 survey respondents, only one opted out.

\subsubsection*{A Word on Generalizability}
As mentioned above, this report is not research, an activity whose goal is the construction through observation of generalizable models of understanding and prediction for observed phenomena.  Most of the data we collected is ``ordinal data,'' and we are largely interested in the proportions and frequencies of responses among all respondents in order to provide a snapshot of the current moment, rather than in, e.g., determining statistical significance between different groups of respondents or between a ``control'' and ``test'' group.  The point is: we believe that our results are important and likely representative of general APS member disposition, even given the small sample size, but we do not claim absolute generalizability across all 50,000 APS members -- we would welcome any opportunity to conduct such a measurement, and hold open the offer of collaboration to APS leadership.  For those interested, we encourage readers unfamiliar with physics education research (PER) techniques, which underpinned our design and analysis of the survey, to consult a few seminal guides on qualitative and quantitative research \cite{ding2012getting,otero2023qualitative}.

\section{Results and Themes}

We present here general themes and specific points of interest from survey results.  We share one note before doing so: given that every single item on the survey was optional, not every respondent chose to answer every item.  We therefore endeavor to share the total number of respondents for a given item or section of questions.  We have opted to share an overview of the quantitative data first for the questions about APS communication and the questions about member needs, then provide additional context from the qualitative responses.  For the Likert-style questions, we generally collapse answers into ``favorable'' and ``unfavorable'' responses, but break individual categories out in a few places where to do so is significant for the purposes of understanding members' responses.  Astute readers may note that, occasionally, percentages for a given item do not sum to 100\%.  The percentages calculated for each response indicate how many respondents answered that specific question out of the total number of respondents who answered any question in the block.  While most respondents answered all questions, a very small number skipped one or more items in a block, leading to response rates that do not sum to 100\% for a given item.  That is, in items where the stated numbers do not add to 100\%, it has been left implied that the remaining percentage of respondents had no answer for that item.

\subsection{Awareness of and Opinions on APS Communication}
Between 93 and 100 of respondents answered the Likert-style questions on APS communications.  We are thrilled first and foremost to report that a majority of respondents to these questions -- between 74 and 80\% -- either agreed or strongly agreed that APS is supporting them, that the society cares about their current needs, that it will continue to be a reliable source of support, or that it is transparent about its policies and practices.  However, we note that this means that up to 26\% of respondents -- one in four -- do \textit{not} agree with these sentiments.  \callout{In particular, 23\% of respondents said they did not agree that APS was currently supporting them and 26\% of respondents did not agree that APS ``will continue to be a reliable source of support in the future.''} These results are summarized in Table \ref{table:apscomms1}.

\begin{wraptable}{r}{7.75cm}
   \includegraphics[width=0.525\textwidth]{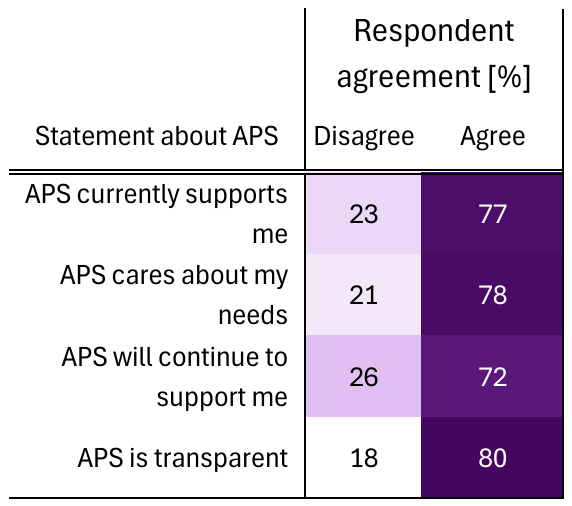}
    \caption{Respondent disposition to questions about perceptions of APS's level of support.  The darker the shade backing the percentage, the higher the proportion of respondents who either agree or disagree.  Full articulations of the statements can be found in Appendix \ref{app:survey}.}
    \label{table:apscomms1}
\end{wraptable}

Similarly, we found that, while a majority of respondents felt that APS is doing a good job of communicating a commitment to uphold the APS values, a willingness to listen to members, a committment to supporting members facing personal challenges, and transparent timely updates on activity, appreciable subsets of respondents -- between 9 and 30\% -- disagreed with these sentiments.  \callout{In particular we note that 24\% of respondents indicated that they did not agree that APS was communicating a willingness to listen to members' needs and interests before acting.}  We also want to highlight in particular that, while \textit{90\%} of respondents did agree that APS is communicating a commitment to uphold APS values, only 68\% agreed that APS was communicating a commitment to supporting APS members themselves in the face of hardship.  This is a substantial difference, and we highlight it to highlight that respondents appear to perceive a difference between the way that APS talks about itself and the way it talks about them.  These results are summarized in Table \ref{table:apscomms2}.

Short-answer responses to this question block may provide additional context; we discuss them below in Section \ref{sec:themes}.  We also  present in Table \ref{table:apsefforts} the results to the question on which APS efforts respondents were aware of at the time of taking the survey.  While these results do not directly address the question of member needs, we do note that only 62\% of respondents indicated awareness of the one explicitly member-focused APS effort listed, whereas 82\% were aware of opportunities to donate to APS.  The member hardship waiver is a great example of person-focused support fom APS and is exactly the kind of effort which the joint task force described above in Section \ref{sec:motivation} could help improve member awareness of (objective {\color{section}(ii)} of the task force) -- we call it out here explicitly in case any readers are experiencing hardship and unaware of this opportunity.

\subsection{Member Needs and Impacts}
The second block of Likert items articulated a variety of actions which an organization like APS could potentially take in service of its members: emergency funds, connection to legal representation, guidance on local advocacy opportunities, etc., and asked how helpful each of these approaches would be for the respondent. 100 respondents answered this block; in this case we have opted not to condense responses into positive/negative categories but provide the dataset in its entirety in Table \ref{table:needs}.  The actions in the leftmost column are listed in shorthand to save space, but their full articulation is in the appended survey (Appendix \ref{app:survey}).

\begin{wraptable}{l}{7.75cm}
    \includegraphics[width=0.525\textwidth]{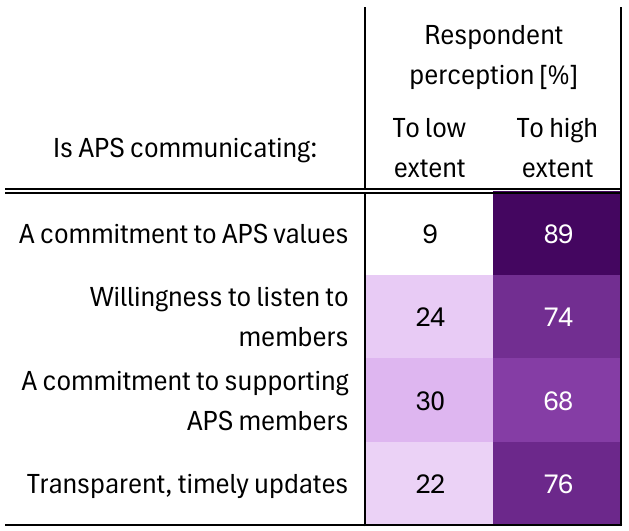}
    \caption{Respondent perception of APS's communication about various topics.  The darker the shade backing the percentage, the higher the proportion of respondents who think APS is communicating the topic to a low extent or high extent.  Full articulation of the topics can be found in Appendix \ref{app:survey}.}
    \label{table:apscomms2}
\end{wraptable}

The items above which garnered a majority positive response (that is, a majority of respondents indicated that the approach would be helpful to them either ``to a great extent'' or ``to some extent'') are: ``Resources for finding programs and career pathways that are robust in the current climate'' (69\%) ``Guidance about how to find local resources related to science advocacy'' (74\%), ``Joining or promoting a consortium of professional societies to collaborate on addressing the current impact on science'' (75\%), and ``Providing space or channels to connect with other APS members on current issues'' (67\%).  47\%, nearly one of every two respondents, said they would benefit from making emergency funds available to members, 38\% said they would benefit from connections to legal representation (note that we asked explicitly about \textit{connections} to legal representation, not APS representing members in court), and 46\% said they would benefit from resources for physicists facing immigration issues such as deportation.  That is to say, every single one of these approaches was desirable for at least four of every ten respondents.  \callout{In particular, we highlight that a majority of respondents indicated strong interest (``to a great extent'') in both coalition-building and in career pathways resources as support avenues, and local advocacy support nearly received a majority of strong support as well -- suggesting that these three avenues may be particularly impactful for future action.}

\begin{wraptable}[15]{r}{7.75cm}
        \includegraphics[width=0.525\textwidth]{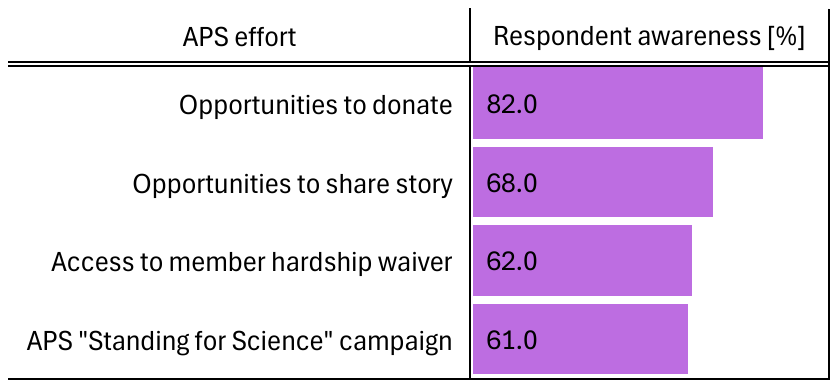}
    \caption{Respondent awareness to various currently-ongoing APS activities and efforts.  The longer the bar, the higher percentage of respondents indicated awareness of the listed activity.  full articulation of the activities can be found in Appendix \ref{app:survey}.}
    \label{table:apsefforts}
\end{wraptable}

Of 64 respondents that answered the bank of yes/no questions about impacts to their professional lives,  47\% said they had been impacted by the recent executive orders, funding cuts, and reductions-in-force, 42\% said that recent events had impacted their approach to international travel, 6\% said they, their students, or their colleagues had been denied entry to the U.S. or detained at the border, and 39\% said that recent events had created immigration or visa issues for them, their students, or their colleagues.  Because these questions are necessarily more personal than the questions from the Likert question banks, we use them as the starting point for presentation of the open-ended questions for each section.

\newpage

\subsection{Themes of Open-Ended Responses}
\label{sec:themes}

\subsubsection{Impacts on Professional Life}

For each of the yes/no questions, there were between 2 and 13 additional open-ended responses, roughly proportional to the number of respondents who indicated that they had indeed experienced the particular impact.  Comments indicated primarily fear for the future -- not just of individuals' research work and livelihoods but of their safety and that of their colleagues.  In particular, responses indicated concern about students, either those currently here on visas or those whose research was threatened, for example by ``blacklists'' like Texas senator Ted Cruz's list of ``woke DEI grants'' \cite{NPR_Cruz_2025}.  \callout{In addition, respondents indicated a general hopelessness and anxiety, a lack of trust in institutions, and a sense of feeling abandoned by both their own institutions and by the country overall.}  A large number of respondents described postponed or canceled travel across the U.S. border, lower interest in attending conferences to which they could not drive, including the APS Global Physics Summit, and overall fear of passport- or identification-related trouble as a function of their identity.

\begin{wraptable}[28]{r}{7.75cm}
    
        \includegraphics[width=0.525\textwidth]{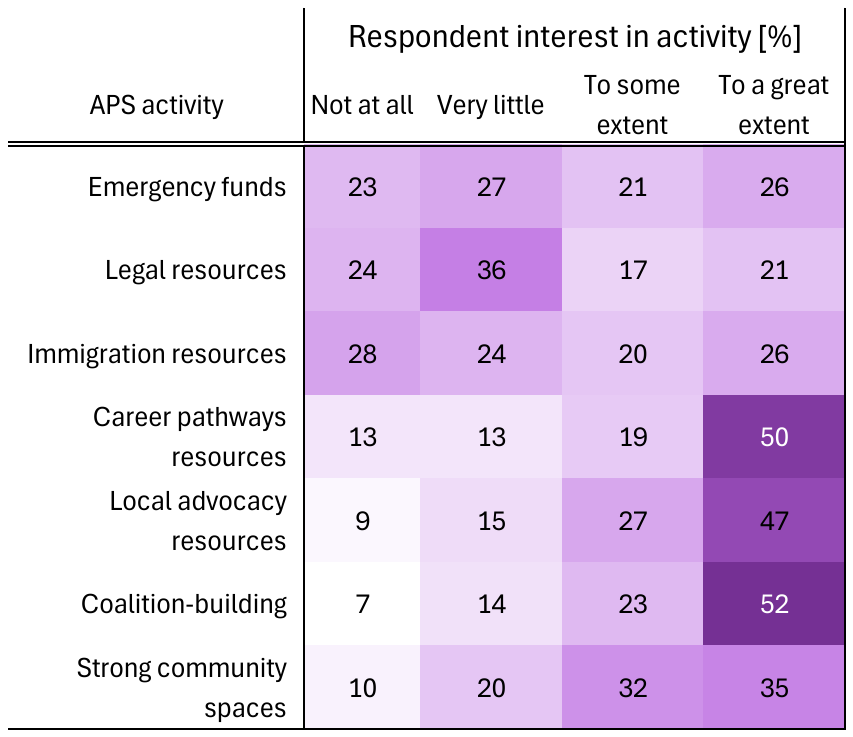}

    \caption{Respondent interest in various potential APS support activities.  The darker the shade backing the percentage, the higher proportion of respondents favoring a particular approach to a great extent, to some extent, very little, or not at all.  Percentages can be summed to condense responses to ''favorable'' or ''unfavorable'' toward a particular activity, but in this case we present the data ``uncompressed'' to highlight actions prompting particularly strong interest.  Full articulations of the proposed activities can be found in Appendix \ref{app:survey}.}
    \label{table:needs}
\end{wraptable}

\subsubsection{Needs and APS Efforts}
27 respondents provided additional comments on the Likert block about their needs and potential APS support efforts.  The most common themes were: desire for funding opportunities, especially for early-career and marginalized groups, as well as for non-R1 institutions; discontent with a perceived APS leadership focus on corporate/executive interests rather than membership needs; and a desire for more explicit support for marginalized groups (e.g., neurodivergent physicists, gender/sex/romantic minorities (GSRM), immigrants, etc.).  Responses named actions by the administration as dangerous and reckless, and articulated additional forms of advocacy such as direct-to-public advocacy as potential avenues of utility in addition to lobbying efforts.  \callout{Multiple responses articulated explicitly that they felt ignored by APS, either as a result of a perceived organizational focus on corporate priorities or as a result of direct denial from the organization in attempts to help vulnerable groups.}

\subsubsection{APS Communication}
29 respondents provided additional comments on types of APS communication that would be helpful for them.  The most common themes were: more info and support for specific groups (early career physicists, marginalized groups, and international students were all named); a stronger communicated commitment to ``leading the resistance'' and to being a standard-bearer as the premier physics professional society in the U.S., and greater centralization of communications to reduce overload and burnout on the part of members.  Multiple comments explicitly named a desire to see APS work together with other science groups.  A number of responses also indicated that communication focused on supporting ``physics'' as an entity and as securing research funding as an end goal misses the mark in the minds of vulnerable APS members for whom funding needs are secondary in importance to threats of deportation, loss of health care, and other personal dangers.  

\subsection{Demographics and Group-Specific Responses}
As mentioned above, we asked a small set of demographic questions for the purposes of gauging what types of physicists were responding to the survey.  64 respondents gave information about their career role: of those, half identified as ``faculty, staff, or scientist,'' 19\% identified as a graduate student, and between 2 and 9\% identified as other roles.  31 respondents answered the question about their career stage, with 54\% of those respondents saying they were tenure-track, 16\% saying they were early career, 16\% saying they were non-tenure track, and a small number identifying in the other categories.  63 respondents gave information about their place of employment; 71\% of those respondents work at a college or university, 10\% at a national lab, and small numbers work at the other categories.  12 respondents gave additional demographic information; these responses tended to reflect positionality in physics both in terms of role (e.g., as a PhD candidate, as a public engagement practitioner) and in terms of minority affiliation.

To see if there were appreciable differences between ``populations'' of respondents, we looked at proportions of responses for the quantitative questions gated on whether respondents were ``early career'' (undergraduates, grad students, and postdocs) or ''late career'' (tenure-track, $<5$ years from retirement, or retired).  Broadly, we found that interests, perceptions, and needs were not dissimilar between groups; however, one interesting distinction did emerge. The 21 respondents who self-identified as ``early career'' were far more likely to be interested in career pathways resources (90\% interest vs. 69\% general interest), but also expressed greater interest than the overall sample group in both local advocacy resources (85\% interest vs. 74\% general interest) and in strong community spaces (81\% interest vs. 67\% general interest).  While both groups expressed 75-80\% interest in coalition building as an APS activity, we did note an interesting distinction, in that, while early career respondents were evenly split between strong interest and moderate interest (about 38\% each), 60\% of the 41 respondents who self-identified as late-career expressed strong interest in coalition building (20\% expressed moderate interest).  That is to say, in our sample, older, while there was broad interest in collaborative, coalition-focused advocacy efforts at APS, it is apparently an appreciably higher priority for older, more experienced physicists.

\section{Discussion and Recommendations}
\label{sec:discussion}

We focus our discussion primarily on the questions of APS communications and members' expressed needs and priorities.  Because the goal of our original proposed task force was to identify member needs and connect APS action to those needs, that lens informs much of our commentary on the results of our survey.  We also include some suggestions for ways in which APS could address concerns arising from the results of the survey in moving forward as a member-led organization.

\subsection{APS Communication}
\callout{We want to be unequivocal in our initial assessment: it is \textit{great} that a majority of respondents think APS is committed to supporting them, and it is great that a huge majority perceive APS to be committed to its values.}  APS probably cannot hope to make all of its 50,000 members happy all of the time, and we do not wish to imply that the Society should be all things to all people.  However, it is indeed troubling that up to one quarter of respondents \textit{do not} perceive APS to be a supportive organization committed to \textit{them.}  \callout{25\% may not seem like a large number of disaffected physicists, but in a 50,000-person organization, the idea that over 12,000 members are not content with current organizational activity is alarming.}

It is also highly concerning that so many more respondents perceive APS to be committed to its on-paper values than it is committed to embodying those values through meeting members' needs -- as mentioned above, approximately 30\% perceive a disconnect between the way APS talks about itself and the way it talks about and to its members.  This disconnect could be related to the expressed perception of APS as driven by corporate/executive interests vs. actual member interests.  While a majority of respondents did report perceiving APS as transparent, additional actions to \textit{demonstrate} transparency could go a long way to increasing the share of members that trust the organization.

As an example, APS performed a member survey in May of 2024 asking membership when the organization should make public statements.  In conversations during and after the GPS town halls, APS leadership highlighted the results of that survey -- the finding that respondents wanted the organization to make statements only when relevant to physics and physics expertise -- but also acknowledged that they had not surveyed membership for its general disposition on speaking out now, after the inauguration of the current administration.  \callout{Such an effort could be a very simple way to check the temperature of members and to ensure that APS's organizational and leadership priorities are aligned with members' priorities.}

We also recommend that APS make it a priority in communications not simply to acknowledge the dire situation presented to science funding but the real, tangible harms done to the \textit{humans} who make up the scientific community, many of whom are dues-paying APS members.  \callout{If over three of every ten dues-paying members do not think the Soceity is committed to supporting them, it raises the question of the purpose of those members' dues.}  We do note that recent member communications from, e.g., APS Government Affairs have noted that, for example, ``rash threats and actions contradict moral values of fairness and meritocracy,'' and that ``international students and scientists should never be used as political tools.'' These are great steps, and we encourage APS to be even more explicit about where these threats are coming from and the impact on APS members specifically.

\subsection{Needs and Impacts: Collaboration and Advocacy}

If our survey results are representative of APS membership, then APS membership is clearly ready for \textit{more} when it comes to advocating for each other and for our field.  Nearly three-quarters of respondents indicated a desire to see more guidance about how to find \textit{local} opportunities for science advocacy, \callout{and the most-desired approach to APS support noted in survey responses was the desire to see APS join a large-scale collaboration between societies.}  As of publication of this report, APS maintains an unlinked page listing seven collaborating organizations ``standing up for science'' along with APS to support science funding \cite{APS_collab_site}.  With the exception of the National Society for Hispanic Physicists, however, every one of these collaborating units is another member society of the American Institute of Physics (AIP).  As of publication, we were unable to find any instances of APS collaborating with other large professional societies such as the AGU, the American Association for the Advancement of Science (AAAS), etc.  To our knowledge, APS is also not listed as a signatory \footnote{We do note that, as APS leadership highlighted at the GPS town halls, the organization did file an amicus brief on behalf of recently-fired federal workers \cite{APS_amicus}. Amicus briefs are ``friend of the court'' opinions from non-parties to a lawsuit that may provide guidance or insight while deciding the case \cite{LE_amicus}. Brief filers do not have a legal stake in the lawsuit. The impact from amicus briefs is not perfectly clear in non-Supreme Court contexts but is generally thought to be positive \cite{kearney1999influence,sungaila2024makes}. The judge's ruling in favor of the federal employees in the suit mentioned in Ref. \cite{APS_amicus} did not mention any amicus briefs as relevant to the decision, and we note as well that AGU was listed as a plaintiff in the lawsuit \cite{afge_lawsuit}.  AGU lists 62,000 members, a substantial portion of which, like APS, are likely federal employees.  We include this context for those who, like us, are not experts as to what it means for an organization to get involved with a lawsuit at various levels.} in any lawsuits challenging the legality of the administration's efforts to dismantle science or as a collaborating organization in large-scale campaigns like the Save NSF campaign \cite{savensf_website}.  

As of the publication of this report, the American Association of Physics Teachers, another AIP organization, has joined with the American Education Research Association (AERA), the American Association of Colleges and Universities (AAC\&U), the Women in Engineering ProActive Network (WEPAN), the American Association of University Professors (AAUP), and United Auto Workers (UAW) in filing a legal challenge to the recent massive NSF grant terminations \cite{democracyforward_lawsuit}. It is in some ways surprising to see that smaller professional organizations are willing to step up to the legal plate and challenge federal attacks when large societies are not.

Greater collaboration, or, cynically, even the \textit{appearance} of greater efforts to collaborate, may help dispel the perception that APS is driven by corporate interests (i.e., those of executive leadership) rather than member needs.  There is a well-known``physicist arrogance'' problem \cite{gibson2003arrogance}, stereotypically represented by the physicst refusing to act collaboratively with even those professionals who could support physics activity, simply because ``physicists can do it all themselves.''  In a situation like the current climate, where every single scientific field is facing an existential threat, it would seem obvious to collaborate as broadly as possible in efforts to stand up for science.  However, many professional societies seem to be conducting independent campaigns with little to no acknowledgement or even awareness of duplication of effort.  We do acknowledge that this problem is not unique to APS; other large professional organizations apart from AGU have also not taken action to build a coalition, leaving the scientific response to federal attacks a piecemeal one.

As of the writing of this report, APS \textit{has} offered a webinar focused on finding avenues for local connection using public engagement techniques \cite{APS_community_webinar}.  This is a great step and we applaud the effort to support members in making direct connections to their community.  From the webinar description, it is unclear to us whether it is an advocacy-focused effort or a public engagement focused effort (or, indeed, both -- our belief is that public engagement is a crucial element of advocacy).  We might recommend that, if this is indeed part of a push to support local advocacy, APS take a more explicit communication tactic about the advocacy angle, similar to the level of explicitness that it uses in drawing members to its lobbying campaign.  While this may seem pedantic, we do note again that 30\% of respondents indicated that they did not perceive APS to be committed to supporting the needs of the actual people that make up its membership.  Person-focused advocacy is essential right now; communicating a person-focused strategic priority is crucial.
 
\subsection{Needs and Impacts: Person-Focused Support}
Although APS communications have taken an increasingly personal timbre since the organization's initial, more circumspect member outreach, the bulk of their apparent actions, and the activities to which they have invited members, still appear largely focused on the strategic goal of maintaining federal funding through lobbying efforts.  But our survey results demonstrate a strong member interest in specific, person-focused avenues of support, whether support for early career members' desire to remain safe while pursuing physics (through, e.g., identification of programs likely to be safe for marginalized students), access to emergency funds, legal resourecs, or immigration support, or even just space to connect more directly with other APS members.  Physicists are human people, and humans evolved for community -- support of a community and support of its people are synonymous.

\callout{In particular, we want to highlight that none of the avenues of support respondents expressed interest in require APS to be the \textit{direct or sole provider} of support.}  Compiling resources for early career physicists does not require APS to find anyone a job.  Pointing members to resources or legal support of which they can avail themselves in the face of deportation does not require APS to represent them.  Starting an emergency fund to which members can contribute does not require APS to spend any of the nearly \$40 million net income it reported in its most recent tax filings \cite{propublica_APS_profile}.  But respondents expressed a deep longing for support that goes beyond opportunities to lobby on behalf of ``physics'' or on behalf of the American Physical Society.  \callout{These support modalities and the interest even their mention generated demonstrates the depth of desires for APS to meet its members basic human needs: not simply their resource needs but the need for connection to others, the need to feel safe in one's environment, including their professional environment, and the need for group solidarity against outside threats.}  Funding is an important element of survival of the physics trade, yes -- and we again laud APS's efforts to stand up for science.  But without standing up for the \textit{scientists} that make up the trade, we fear that APS risks saving a version of the field that will, by that point, have regressed a century or more, keeping only the least vulnerable among us who tend to be wealthy, white, male, well-connected, etc.  We encourage APS to consider the question of \textit{on behalf of whom} they are standing up for science.  Is it for the most vulnerable among us?  Or is it for the status quo?

\subsection{Needs and Impacts: Leading the Resistance?}

As Ref. \cite{APS_collab_site} demonstrates, the organization is eager to call members to action and highlight collaborative efforts to meet challenges, but it appears less willing to explicitly set itself \textit{against} any of these challenges or name them for the existential threats they are, and has stopped short of calling out actors and entities whose attacks are causing the challenges in the first place.  For example, on the same day on which news of the AAPT/AERA coalition's lawsuit against NSF cuts dropped, the APS President and CEO sent out an email in which they ``recognize that recent events -- including arrests, visa revocations, and severe delays in processing -- are causing many to feel targeted and threatened,'' and claiming that ``[t]hat's why APS is standing with you, pushing back on policies that contravene our values.''  But nowhere in the letter does APS leadership identify the source or nature of the actions \textit{causing} visas to be revoked, \textit{causing} arrests, \textit{causing} physicists to feel targeted and threatened -- as though these harmful actions merely appeared out of the luminiferous \ae ther  To its credit, APS does list in that letter several actions that could indeed positively impact members, including meeting with members of Congress to encourage fair visa policies, preparing legal briefs, and communicating with policymakers the value of international science collaboration.  But we also note that nowhere does APS provide links for members to learn more about any of the actions mentioned or their actual, tangible impact on members, nor does it provide opportunities to partner on those actions or even to weigh in on them at all.

As the pre-eminent physics professional society in the U.S., APS has appreciable ``moral'' authority in the physics sphere in exercising what economists would call ``moral suasion'' \cite{Investopedia_moral_suasion}.  \callout{The actions to which APS calls its members and the actions in which the organization itself engages demonstrate to politicians, to the public, and to APS members where the organization's priorities lie, and where the organization thinks its members priorities \textit{should} lie.}  If the organization is more interested in submitted non-party briefs to existing lawsuits than it is in becoming a party to lawsuits on behalf of its members, and it does nothing to explain that stance or engage with members on its decisions, what conclusion should members draw about the organization's commitment to their continued professional existence relative to its own?  Given that a significant number of responses to the survey's open-ended questions touched on the perception that APS is less committed to member needs than its own corporate interests, a perspective apparently shared by 30\% of respondents with opinions on APS communications, we suggest that APS could prevent an ongoing erosion of trust by being a standard-bearer for resistance not only to the practical harms of funding cuts, but the existential threat of attempts to exercise authoritarian control over science.

APS values include ``Equity, Diversity, and Respect'' as well as ``Trust, Integrity, and Ethical Conduct'' \cite{APS_values}. To-date, to our knowledge, no APS communications focused on federal cuts from the administration have expressed solidarity with, or even acknowledged that outsized impact of cuts on, already-marginalized groups or even certain subfields of physics.  \callout{While communications focused on the funding may preserve working relationships with policy makers and keep APS as an organization in the conversation with those whose values exclude equity, diversity, or respect for marginalized groups, a focus on keeping relationships productive with those perpetuating harm against APS members unfortunately sends the message -- already received too often -- that APS does not value keeping relationships productive with its own members and only desires their incorporation in times when their existence isn't a burden to the organization's political ambitions.}  We exhort APS to act with integrity when it comes to its members, so that members continue to trust that the organization is acting ethically.

\section{Conclusions: Advocacy for Physics?  Or for Physicists?}
The June 18 communiqu\'e mentioned above includes the following caveat about APS's activities: ``We cannot address everything, everywhere, all at once, so we are focusing our efforts on areas where we can have impact.''  This statement, seemingly referencing the Academy Award Best-Picture-winning film of the same name, seems defensible at first -- who wouldn't want their organization's efforts to be focused where it can have impact? -- but it belies a dangerous assumption, namely that the ways in which organizational leadership defines ``impact'' are more to be trusted than the organization members' definitions of impact and priorities for achieving that impact.  Coupled with the organization's lack of actual knowledge about member needs and interests, as well as our experience with leadership's apparent resistance to even \textit{learning} about member needs, the implication that no member interests or priorities could be more impactful than the actions APS is already taking is concerning.

For example, as mentioned above, several professional societies have joined a lawsuit to challenge the NSF's termination of grant funding.  APS's claim of focusing action ``where we can have impact'' would seem to imply that the organization considers filing a brief on behalf of this lawsuit, which very obviously concerns a large share of APS members, to be \textit{more} impactful than actually joining the lawsuit on their behalf.  Such an argument seems ludicrous to us, and the rhetoric used to make that argument seems to rely on the condescending assumption that ``APS knows best'' and that members don't need to think critically about how their professional society operates.  Furthermore, are members to understand that, since APS is ``choosing to focus our efforts on areas where we can have impact,'' the organization does not perceive there to be benefit or impact \textit{at all} in any number of member-focused actions which it has not already undertaken of its own volition, or in even listening to members to learn if such actions exist?  Clearly, a significant portion of APS membership thinks it would be impactful to see APS collaborate more broadly with other large professional societies.  Clearly, there is substantial interest in member-focused activity like emergency funds, which, na\"ively, seem like an almost trivial idea to implement, or in creating member spaces to discuss the current climate.  \callout{Our survey results suggest that there is a substantial portion of physicists who don't just feel abandoned by their country: they feel abandoned by APS itself in the fight to save physics.}  We therefore exhort the organization, if it truly wishes to lead the resistance against the current attacks on the scientific enterprise, to prioritize the people who make up that enterprise at \textit{least} as much as it prioritizes the funding that makes the effort possible for \textit{some} of us.

Every one of the individuals who have contributed to this report is an APS member; we want the organization to succeed.  \callout{But to succeed in the midst of these existential threats, the organization needs to acknowledge not only the impact of its actions, but the impact of its} \textit{ \color{section} inactions,} the cost to human capital and human flourishing when refusal to talk to members, listen to members, work with members, and act on behalf of \textit{members} causes one in four members to perceive that the organization does not stand for them.  We exhort APS not to create its own ivory tower, not to embody the worst elements of physics culture in an attempt to merely survive the current administration or, worse, to gain power through cooperation with it, but to express loudly and unequivocally that \textit{organizational success looks like member success.}  It is abundantly clear that federal attacks on science are not equally distributed and are not the result of ignorance of funding agents or misguided attempts at exercising oversight; they are calculated and intentional attempts to bring the scientific population under control, remove its ability to influence society at large, and ensure that it is used only in ways desirable to the administration.

\callout{Any attempt at preserving funding, even if it retains funding for a subgroup of physicists, cannot be called success if it does not result in greater flourishing for the entire scientific community, especially those members most directly and disproportionately affected by these attacks.}  Throwing APS's considerable weight behind any number of grassroots efforts like Save NSF, behind a collaboration with other societies, or behind lawsuits on behalf of its members -- such action would likely turbocharge these efforts and make success -- \textit{true success,} for all physicists -- that much more likely.  If we as scientists only care about our own funding, our own laboratory, our own projects, we risk losing the opportunity to stand up for and protect the foundations on which they are built.

Finally, we want to acknowledge that, indeed, the American Physical Society, a professional organization with a specific focus on physics and ostensibly run by physicists, cannot address everything, everywhere, all at once.  As a member organization, APS is only as strong as its members in terms of identifying and addressing issues of import.  We therefore exhort our fellow APS members not to give in to the cynicism and hopelessness threatening us in this moment, but to step up and act on behalf of one another directly, publicly, and loudly -- and to be direct, public, and loud with the changes we want to see both in society and in our professional organization.  \textit{\color{subsection}We} \callout{are APS: the organization was founded for and by working physicists, and we exhort APS members to use their rights to speech to speak up and speak out when they see problems, both in society and in our organization, whether those problems are attacks on physics funding, attacks on Black physicists, threats to our early career researchers, deportations of the country's brightest minds, threats to academic freedom, or threats to queer physicists.}  We do not need to wait for an institution to give us permission to care for one another, or to shape it how we want.  We encourage anyone reading this report who finds themselves disheartened to view this as an invitation to participate in the shaping of our Society and of American society at large.

\subsubsection{Final Words}
In the film \textit{Everything, Everywhere, All At Once,} referenced in APS's June 18th communiqu\'e, the protagonist, Evelyn Wang, learns of the existence of an infinite number of universes and learns that they are all threatened by Evelyn's daughter Joy, whose over-exposure to the apparent meaninglessness of the multiverse has caused her to sink into a deep nihilism and create an ``Everything Bagel'' that not only \textit{contains} everything -- hopes, dreams, sesame, salt -- but \textit{consumes} everything as a multiversal singularity that threatens Evelyn's own reality and the reality of countless other versions of herself.  In like manner, the United States finds itself under threat from a ``political singularity,'' a force of social nihilism arguing that the only thing that matters is the accumulation and use of power.  For scientists, whose superpower, according to APS President John Doyle, is ``telling the truth,'' this force must be seen for the existential threat that it is.  If power is indeed the only thing that matters, then it indeed makes sense for APS or any organization to avoid taking a moral stance in order to keep relationships productive with those perpetuating harm against its own members, especially if doing so leads to additional power or resources for the organization or those it favors.  But of course this course of action would also betray the same cynicism, the same nihilism demonstrated by Joy in creating the Everything Bagel, and is fundamentally incompatible with APS's stated values.

Throughout the film, Evelyn is told over and over that the only way she can save the multiverse is by drawing on the power of her alternate selves and destroying her daughter.  However, after confronting the truth of Joy's words and the apparent meaninglessness of reality, she nevertheless chooses to \textit{create} meaning, for herself and her alternate selves, by adopting a fundamental posture of empathy toward herself and others.  By persisting in her empathy in the face of cynicism, by choosing to embrace and advocate for her family and her daughter even in situations where to do so doesn't benefit her or even ``impact'' the multiverse at large, Evelyn is able to reconcile with Joy and prevent the destruction of her reality.  We agree with APS: we do not expect the organization, or any of its members, to do everything, to be everywhere, or to do it all at once.  But, as in \textit{Everything, Everywhere, All At Once}, each decision made \textit{today} compounds to create myriad new realities \textit{tomorrow.}  We are, today, all of us, in an unprecedented moment of U.S. scientific and cultural history.  It is incumbent upon us -- all the more so for those of us with organizational power -- to make the most of these ``few specks of time when any of this actually makes any sense'' by choosing empathy for one another and especially for the most vulnerable among us, and by rejecting the lure of the cynical pursuit of power.  One way that the American Physical Society can choose empathy is by refocusing its advocacy efforts on \textit{physicists,} not just on physics, and demonstrate its willingness to take a stand on behalf of its most vulnerable members. 

\section*{Acknowledgements}
The authors of this report are members of the Physics Advocacy Collaboration, a grassroots group of APS members working to advocate on behalf of physicists as people.  We are happy to discuss the report with any interested readers and can be reached at phys\_advocacy@proton.me.  While we claim shared authorship over the report, naming one author was required for the purposes of posting to arXiv, so we have followed the convention for high-energy experiment collaborations.  We would like to express sincere and hearty thanks to GPER, FPS, FOEP, FGSA, FECS, and FEd as units who provided valuable insight and feedback in the design and implementation of this project.

\subsubsection*{Note on Bibliography}
Wherever possible, we have tried to provide references that are ``robust'' against changes over time, while recognizing that the political landscape today often results in agents changing or removing information in an attempt to control the awareness of the general public.  We also note that several references may ask readers to register an account with the publication in order to read.  In both cases, it is possible that information has been registered with the Internet Archive at \url{archive.ph}; in such a case, searching for the URL of the desired web page may lead to an ``archived'' version that has been independently preserved.

%%\bibliography{report}

\printbibliography

\appendix
\newpage

\section{Survey Text}
\label{app:survey}
Appended here, we include the text of the survey.



\section*{Supplementary material}

**Optional: Personal Experiences (Click to expand)** 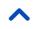

# Thank you for your input!  The following questions are optional.

The following questions are intended to help us better understand how the APS community has been impacted by the current climate.  They are completely optional but will help us gather stories of how physicists have been impacted.

**Have you or your work been impacted by the recent executive orders, funding cuts, RIFs, etc.?**

○ Yes

○ No

**Have recent events impacted your approach to international travel? (e.g., are you limiting your own travel?)**

○ Yes

○ No

**Have you, your students, or your colleagues been denied entry to the U.S. or detained at the border?**

○ Yes

○ No

**Have recent events created immigration/visa issues for you, your students, or your colleagues?**

○ Yes

○ No

**Please share any other ways you or your work has been impacted by the recent executive orders, funding uncertainty, and general socio-political climate.**

As a reminder, these questions are completely optional and your story will not be used to identify you.

[                                                                                     ]

Optional: Demographics (Click to expand) 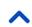

## Finally, some optional demographic questions.

We want to make sure that APS is listening to all of its members. The following questions are again completely optional; their only purpose is to get a sense of whether the needs of different groups of APS members are being met.

**Which of the following best describes your career?**

○ Undergraduate student

○ Graduate student

○ Postdoctoral researcher

○ Faculty, staff, or scientist

○ K-12 educator

- ○ Within 5 years of retirement
- ○ Retired
- ○ Other

**Which of the following best desccribes your place of employment?**

- ○ College/university
- ○ National lab
- ○ For-profit company
- ○ Non-profit organization
- ○ K12 school
- ○ Museum or other cultural institution
- ○ Federal or state agency
- ○ Other

**Please share any other demographic information you'd like us to know.**

# Thank you for your help!

We are considering compiling AGGREGATE results from the survey (i.e., percentages, trends, themes, but NO individual responses) into a report that can be easily shared with APS leadership or published on the arXiv. Do you want your data included in such a report, if it happens?
Again, NO individual responses would be included in this report and NO identifying information will be shared.

- ◯ Yes, please include my results in any such report
- ◯ No, please exclude my results from any such report

We would love to have a strong, grassroots effort to ensure that APS is meeting members' needs. Would you be interested in joining an informal member group whose goal is to explore ways that APS as a professional society can provide support to members according to their current needs?

- ◯ Yes
- ◯ No

Submit

Never share passwords in Fillout forms. Report malicious form



\end{document}